\documentclass[twocolumn,showpacs,a4paper]{revtex4}
\usepackage{graphicx,psfrag}
\begin{document}

\title{Impact of van der Waals forces on the classical shuttle instability}

\author{T. Nord$^1$}
\email[]{nord@fy.chalmers.se}
\author{A. Isacsson$^{2}$}
\affiliation{$^1$Department of Applied Physics, Chalmers
University of
  Technology \\ and G\"oteborg University, SE-412 96 G\"oteborg,
  Sweden\\
  $^2$ Department of Physics, Yale University, P.O. Box 208120, New Haven, CT 06520-8120, USA}
\date{\today}

\begin{abstract}
  The effects of including the van der Waals interaction in the
  modelling of the single electron shuttle have been investigated
  numerically.  It is demonstrated that the relative strength of the
  vdW-forces and the elastic restoring forces determine the
  characteristics of the shuttle instability. In the case of weak
  elastic forces and low voltages the grain is trapped close to one
  lead, and this trapping can be overcome by Coulomb forces by
  applying a bias voltage $V$ larger than a threshold voltage $V_{\rm
    u}$. This allows for grain motion leading to an increase in
  current by several orders of magnitude above the transition voltage
  $V_{\rm u}$. Associated with the process is also hysteresis in the
  I-V characteristics.
\end{abstract}


\maketitle


\section{Introduction}
Nanoelectromechanical systems (NEMS) have seen a great upswing in
interest over the past few years. NEMS show great promise for
applications in areas such as charge detection, sensing, transistors
and memory
devices~\cite{a:01_roukes,a:99_kim,a:00_blick,a:00_rueckes,a:03_kinaret}.
In many NEMS devices the interplay between electrical and mechanical
degrees of freedom is crucial for the transport of charge through the
system. This is due to the Coulomb forces exerted by the charge
carriers which may be strong enough to cause mechanical deformations
of the systems. One system where such interplay drastically changes
the charge transport properties is the single electron shuttle
system~\cite{a:98_gorelik}. There a metallic grain is embedded in an
elastic material between two bulk leads. Since current through that
system is accompanied by charging of the grain, interplay between the
Coulomb forces and the mechanical degrees of freedom can lead to self
oscillations of the grain and electrons can be seen as shuttled across
the gap between the leads. Experimental shuttle related work has been
reported on cantilevers that work as flexible tunneling
contacts~\cite{a:98_erbe,a:01_erbe,a:02_scheible}, single-molecule
transistors~\cite{a:00_park}, colloidal particle
systems~\cite{a:02_nagano}, and macroscopic shuttle
systems~\cite{a:99_tuominen}. Theoretical work has included different
aspects of classical shuttle
systems~\cite{a:98_gorelik,a:98_isacsson,a:01_isacsson,a:01_nishiguchi,a:02_nord},
some noise and accuracy
considerations~\cite{a:99_weiss,a:02_nishiguchi}, different aspects of
quantum mechanical shuttle
models~\cite{a:01_boese,a:02_fedorets1,a:02_armour,a:02_fedorets2,a:03_novotny,a:03_mccarthy},
and shuttling of Cooper pairs in a superconducting shuttle
system~\cite{a:01_gorelik,a:02_isacsson}. For a recent review, see
Ref.~\cite{a:03_shekhter}.

When down-scaling mechanical systems into the nanometer size range,
the importance of short range surface forces, such as van der
Waals(vdW) forces, adhesion forces and Casimir forces, increase
causing problems with for instance stiction. To make realistic models
of NEMS and to design NEMS-devices, or to make optimizations of device
designs, it is thus necessary to take such forces into account.  In
this work we report on a numerical investigation of the effects of
vdW-forces on the shuttle transport mechanism. The main result is that
a new voltage scale comes into play, above which a grain trapped by
the van der Waals force will be set loose due to the Coulomb
interactions between the grain and the substrate. Associated with this
process is a large increase in current as well as hysteretic behavior.
\begin{figure}
  \begin{flushleft} (a) \\ \end{flushleft}
  \psfrag{M}[]{\hspace*{2mm}\raisebox{5mm}{\large{$\mathbf{M}$}}}
  \psfrag{VL}[]{\hspace*{4mm}\raisebox{2mm}{\Large{$\mathbf{V_L}$}}}
  \psfrag{VR}[]{\hspace*{2mm}\raisebox{2mm}{\Large{$\mathbf{V_R}$}}}
  \psfrag{VG}[]{\hspace*{3mm}\raisebox{0mm}{\large{$\mathbf{V_G}$}}}
  \psfrag{KL}[]{\hspace*{4mm}\raisebox{0mm}{}}
  \psfrag{KR}[]{\hspace*{1mm}\raisebox{0mm}{}}
  \psfrag{X}[]{\hspace*{-5mm}\raisebox{1mm}{\large{$\mathbf{X}$}}}
  \psfrag{L}[]{\hspace*{0mm}\raisebox{3mm}{\large{$\mathbf{L}$}}}
  \psfrag{r}[]{\hspace*{1mm}\raisebox{2mm}{\large{$\mathbf{r}$}}}
  \includegraphics[width=6cm]{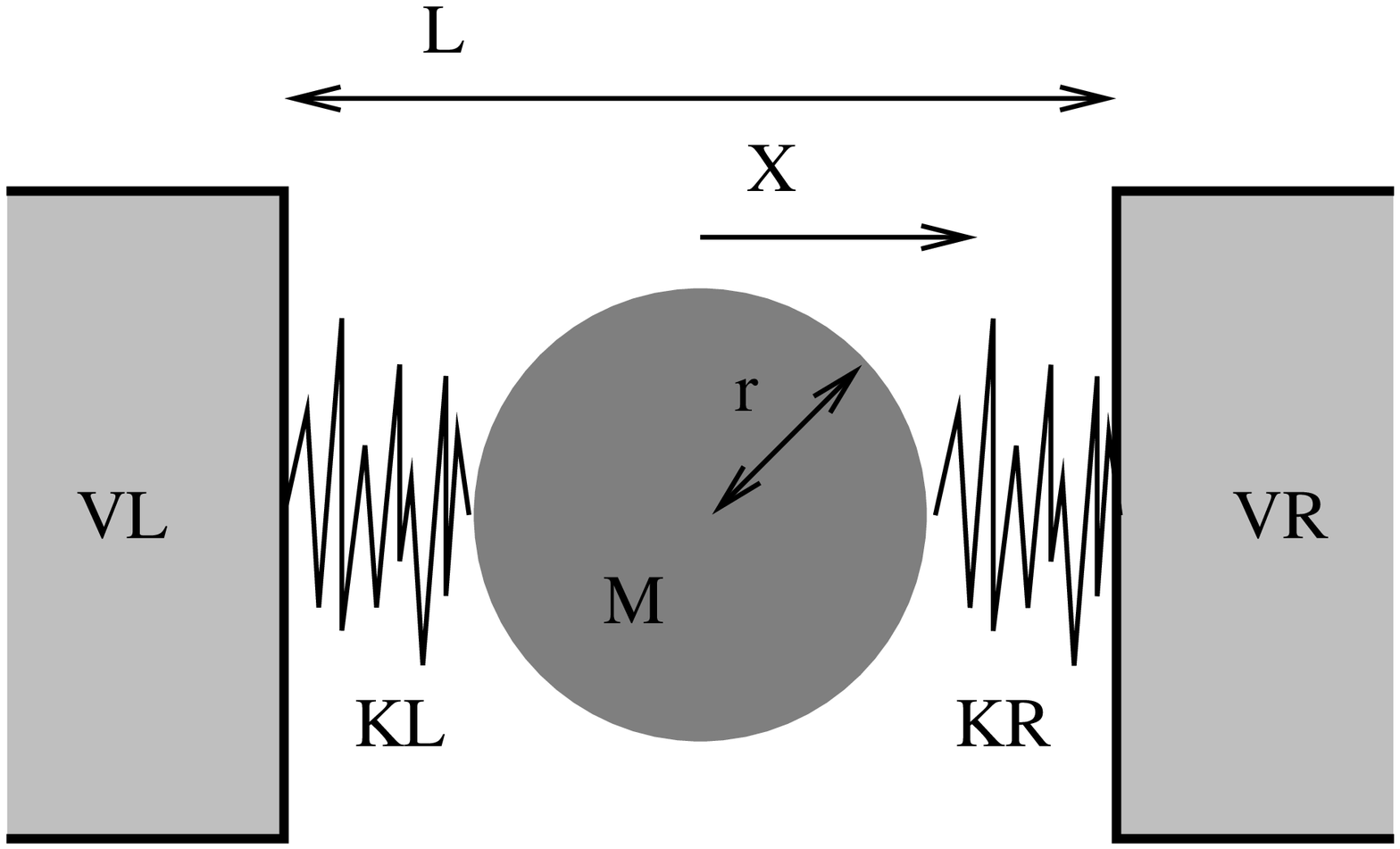}\\
  \begin{flushleft} (b) \\ \end{flushleft}
  \psfrag{RLE}[]{\hspace*{6mm}\raisebox{3mm}{$\mathbf{R_L}$}}
  \psfrag{RRE}[]{\hspace*{3mm}\raisebox{3mm}{$\mathbf{R_R}$}}
  \psfrag{CLE}[]{\hspace*{5mm}\raisebox{3mm}{$\mathbf{C_L}$}}
  \psfrag{CRE}[]{\hspace*{2mm}\raisebox{3mm}{$\mathbf{C_R}$}}
  \psfrag{VLE}[]{\hspace*{5mm}\raisebox{3mm}{$\mathbf{V_L}$}}
  \psfrag{VRE}[]{\hspace*{5mm}\raisebox{2mm}{$\mathbf{V_R}$}}
  \includegraphics[width=6cm]{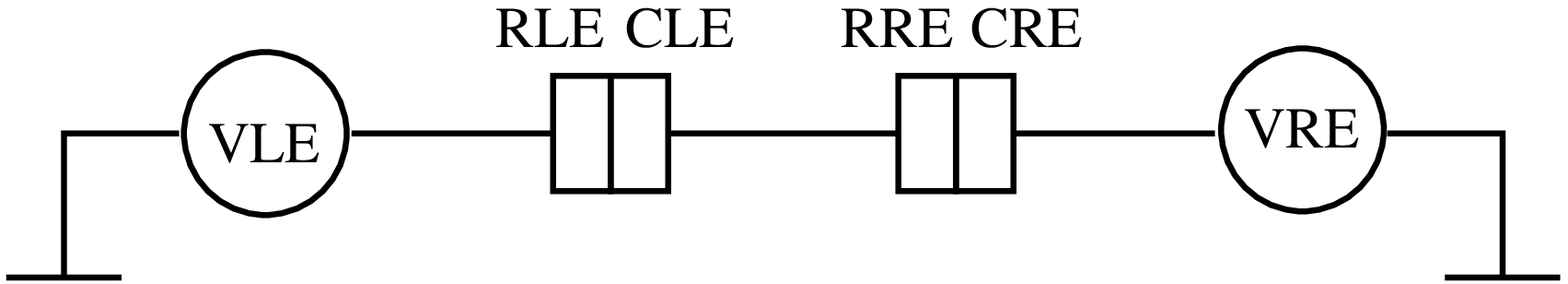}
  \vspace{5 mm}
  \caption{ \label{fig:model_system}Single electron shuttle. (a)
    A metallic grain of mass $M$ and radius $r$ placed between two
    leads separated by a distance $L$.  The displacement of the
    grain from the center of the system is labelled $X$. The grain
    is connected to the leads via insulating elastic ligands. (b)
    Equivalent circuit of the system. The tunneling resistances and
    capacitances of the left and right junctions are $R_L$, $R_R$,
    $C_L$, and $C_R$. The leads are biased to the potentials $V_L$
    and $V_R$.}
\end{figure}

\section{Model System}
To be specific, we will here consider the impact of vdW-forces on a
mechanically soft Coulomb blockade double junction. A schematic
picture of this system is shown in \mbox{Fig.
  \ref{fig:model_system}a}. A metallic grain of mass $M$ and radius
$r$ is placed between two conductors (leads), separated by a distance
$L$, via elastic, insulating materials (ligands). When a bias voltage
$V\equiv (V_L-V_R)$ is applied between the leads, electron transport
through the system occur by sequential, incoherent, quantum mechanical
tunneling of electrons between the leads and the grain.  The system
can, however, also lower its electrostatic energy by altering the
grain position $X$. If the resulting electrostatic force is comparable
to the rigidity of the insulating materials, grain displacement will
occur.  Thus, in this system, both electron transport and grain motion
has to be taken into account.

\subsection{Electron transport}
In the notion of the equivalent circuit of \mbox{Fig.
  \ref{fig:model_system}b} the electrical properties of the system are
characterized by resistances and capacitances $R_L$, $R_R$, $C_L$, and
$C_R$ respectively. It is important that both the tunnel resistances
as well as the capacitances depend on the grain position $X$. The
resistances depend exponentially on $X$ as
$R_{L,R}(X)=R_0^{L,R}\exp(\pm X/\lambda)$ where $\lambda$ is the
characteristic length scale for tunneling. The position dependence of
the capacitances are given by
$$
C_{L,R}(X) = \frac{C_0^{L,R}}{1 \pm \frac{X}{A_{L,R}}},
$$
where $C_0^{L,R}$ are the capacitances of the left and right
junctions when the grain is at $X=0$. The coefficients $A_{L,R}$ are
typical capacitance length scales for the left and right tunnel
junction respectively.

Provided that the charging energy of the grain is larger than the
scale of quantum fluctuations for all positions of the grain, only
sequential tunneling need to be considered and electrons will be
localized in the leads or on the grain. This means $E_C \gg
\hbar/R(X)C(X)$, where $E_C = e^2/2C$.  Tunneling is then well
described by the ``orthodox'' theory of Coulomb
blockade~\cite{a:75_kulik} from which it follows that the rates at
which electrons tunnel through the left and right junctions are
\begin{equation}
\Gamma_{L,R}^\pm(X,V,Q) = \frac{\Delta
G_{L,R}^\pm(X,V,Q)}{e^2R_{L,R}(X)} \frac{1}{1 -
  e^{-\beta\Delta G_{L,R}^\pm(X,V,Q)}
}.\label{eq:rates}
\end{equation}
Here $\beta$ is the inverse temperature and $\Delta
G_{L,R}^\pm(X,V,Q)$ is the decrease of free energy as the event
$(Q,Q_{L,R}) \to (Q \pm e, Q_{L,R} \mp e)$ occurs. The charges
$Q_{L,R}$ and $Q$ denote the charges accumulated on the left and right
leads and the excess charge on the grain. The free energy decrease
$\Delta G_{L,R}^\pm(X,V,Q)$ can be expressed using the equivalent
circuit model of \mbox{Fig.
  \ref{fig:model_system}b}~\cite{a:averin_91}.

\subsection{Grain motion}
\label{sec:forces} Considering the grain motion to be classical
and one dimensional we have Newton's equation $M\ddot{X}=F_{\rm
  Ext.}(X,V,Q)$ for the grain displacement. Whereas in
Ref.~\cite{a:98_gorelik} the external force $F_{\rm Ext.}(X,V,Q)$
acting on the grain was taken to be that of a simple damped harmonic
oscillator driven by a term linear in $Q$, we are here considering a
more realistic model including an elastic force $F_k(X)$, a
dissipative force, $F_{\gamma}(\dot{X})$, an electric force
$F_{\epsilon}(X,Q,V)$ and a vdW force $F_{vdW}(X)$ yielding the
equation of motion
\begin{equation}
M\ddot{X} = F_k(X) + F_{\gamma}(\dot{X}) + F_{\epsilon}(X,Q,V) +
F_{vdW}(X).\label{eq:Newton}
\end{equation}

The elastic force, $F_k(X)$, has its origin in the compressible
material between the grain and the leads. Since the grain displacement
resulting from the electrostatic force may lie outside the range of
validity of the harmonic approximation, the nonlinearity of the
elastic force has to be accounted for. In order to achieve this we
have used a phenomenological potential that acts as a harmonic well
with a usual spring force constant $k$ for small displacements but
that diverges as a Lennard-Jones potential (12th power) at the
positions $X_L$ and $X_R$. The force is thus
$$
F_k(X) = \frac{a}{(X+X_L)^{13}} + \frac{b}{(X-X_R)^{13}},
$$
where
\begin{eqnarray}
  a &=& - \frac{k}{13} \frac{(X_0-X_R)(X_0+X_L)^{14}}{X_L+X_R}
  \nonumber \\
  b &=& \frac{k}{13} \frac{(X_0+X_L)(X_0-X_R)^{14}}{X_L+X_R}.
  \nonumber
\end{eqnarray}
This elastic force is accompanied by a dissipative force,
$F_\gamma(\dot{X})$ which is, as in
\mbox{\cite{a:98_gorelik,a:02_nishiguchi}}, phenomenologically modeled
as a viscous damping term
$$
F_\gamma(\dot{X}) = -\gamma \dot{X}.
$$
The electrostatic force, $F_{\epsilon}(X,Q,V)$, is given by the
derivative of the electrostatic free energy $G(X,V,Q)$
\begin{eqnarray}
  F_\epsilon(X,V,Q) &\equiv& -\frac{d}{dX}G(X,V,Q) \nonumber \\
  &=&\frac{1}{2}\frac{dQ_L}{dX}V_L + \frac{1}{2}\frac{dQ_R}{dX}V_R -
  \frac{1}{2}Q\frac{dV}{dX},\nonumber
\end{eqnarray}
and $Q$ and $V_{L,R}$ can be calculated from the equivalent circuit of
\mbox{Fig.  \ref{fig:model_system}b}.
\begin{figure}
  \psfrag{yp3}[]{\hspace*{1mm}\raisebox{2mm}{{$\mathbf{3}$}}}
  \psfrag{yp2}[]{\hspace*{1mm}\raisebox{2mm}{{$\mathbf{2}$}}}
  \psfrag{yp1}[]{\hspace*{1mm}\raisebox{2mm}{{$\mathbf{1}$}}}
  \psfrag{yn0}[]{\hspace*{1mm}\raisebox{2mm}{{$\mathbf{0}$}}}
  \psfrag{ym3}[]{\hspace*{0mm}\raisebox{2mm}{{$\mathbf{-3}$}}}
  \psfrag{ym2}[]{\hspace*{0mm}\raisebox{2mm}{{$\mathbf{-2}$}}}
  \psfrag{ym1}[]{\hspace*{0mm}\raisebox{2mm}{{$\mathbf{-1}$}}}
  \psfrag{exponent }[]{\hspace*{5mm}\raisebox{3mm}{\small{x
        $\mathbf{10^{-21}}$}}}
  \psfrag{ylabel}[]{\hspace*{2mm}\raisebox{8mm}{{\bf Pot.}
  \hspace*{2mm} {\bf (J)}}}
  \psfrag{xp}[]{\hspace*{2mm}\raisebox{-4mm}{{$\mathbf{2}$}}}
  \psfrag{xn}[]{\hspace*{2mm}\raisebox{-4mm}{{$\mathbf{0}$}}}
  \psfrag{xm}[]{\hspace*{2mm}\raisebox{-4mm}{{$\mathbf{-2}$}}}
  \psfrag{xlabel}[]{\hspace*{12mm}\raisebox{-11mm}{{\bf X}
      \hspace*{2mm} {\bf (nm)}}}
  \includegraphics[width=8cm]{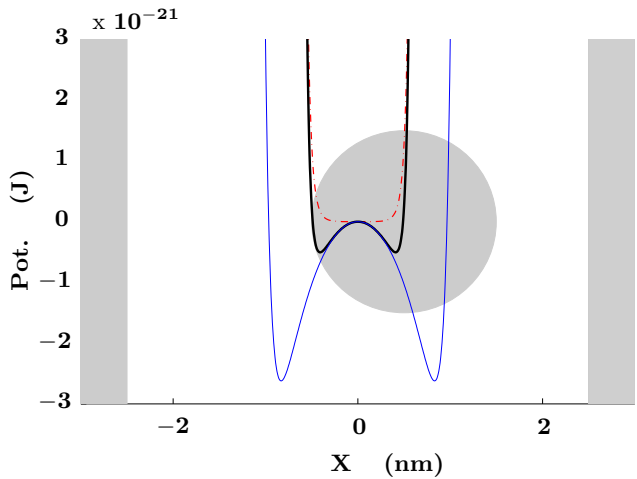}
  \vspace*{5mm}
  \caption{Elastic potential and vdW-potential plotted together
    with their sum. The whole graph is superimposed on a schematic
    figure of the system showing the grain and the leads. The dashed
    curve (red) shows the elastic potential arising from the the
    molecular links while the thin solid line (blue) is the vdW-force
    curve. The total potential is shown as a thick line (black).
    \label{fig:system2}}
\end{figure}

The van der Waals force, $F_{vdW}(X)$, between the grain and the leads
can be derived from a Lennard-Jones interaction potential for the
individual atoms~\cite{a:israelachvili_85} in the leads and the grain
respectively. For a geometry with a metallic sphere close to a flat
substrate one finds,
\begin{eqnarray}
  F_{vdW} &=& F_{attr} + F_{rep} \nonumber \\
  &=& \frac{H_{attr} r}{6}
  \left( \frac{1}{ \left( \frac{L}{2}-r-X \right)^2}
    - \frac{1}{ \left( \frac{L}{2}-r+X \right)^2} \right)
  \nonumber \\
  & & - \frac{H_{rep} r}{180}
  \left( \frac{1}{ \left( \frac{L}{2}-r-X \right)^8}
    - \frac{1}{ \left( \frac{L}{2}-r+X \right)^8 } \right) \nonumber,
\end{eqnarray}
where $H_{attr,rep}$ are the Hamaker constants for the attractive and
repulsive parts of the force. This force has not been considered in
this context before.

Equation~(\ref{eq:Newton}) together with equation~(\ref{eq:rates}) for
the charge transfer rates describe the dynamics of the system.  The
complicated form of the potentials and forces in the present model
makes a detailed analytical treatment cumbersome. However, there is no
problem in treating the equations numerically and the ensuing results
are easily interpreted. Below we introduce a model system with a
specific set of parameter values and investigate numerically the
effect of incorporating adhesive surface forces on the system
dynamics. The numerical investigation has been performed by doing
Monte-Carlo style simulations incorporating a 4th-order Runge-Kutta
routine when solving for the grain dynamics.

\section{Results}
\begin{figure}
  \psfrag{ym5}[]{\hspace*{2mm}\raisebox{2mm}{{$\mathbf{-4}$}}}
  \psfrag{y0}[]{\hspace*{2mm}\raisebox{2mm}{{$\mathbf{0}$}}}
  \psfrag{yp5}[]{\hspace*{2mm}\raisebox{2mm}{{$\mathbf{4}$}}}
  \psfrag{yp10}[]{\hspace*{2mm}\raisebox{2mm}{{$\mathbf{8}$}}}
  \psfrag{yexp}[]{\hspace*{15mm}\raisebox{3mm}{\small{x $\mathbf{10^{-21}}$}}}
  \psfrag{yfiglabel}[]{\hspace*{6mm}\raisebox{8mm}{{\bf Pot.} \hspace*{2mm} {\bf (J)}}}
  \psfrag{xm4}[]{\hspace*{3mm}\raisebox{0mm}{{$\mathbf{-4}$}}}
  \psfrag{x0}[]{\hspace*{3mm}\raisebox{0mm}{{$\mathbf{0}$}}}
  \psfrag{xp4}[]{\hspace*{3mm}\raisebox{0mm}{{$\mathbf{4}$}}}
  \psfrag{xexp}[]{\hspace*{10mm}\raisebox{0mm}{\mbox{}}}
  \psfrag{xfiglabel}[]{\hspace*{12mm}\raisebox{-7mm}{{\bf X} \hspace*{2mm} {\bf (\AA)}}}
  \includegraphics[scale=0.3]{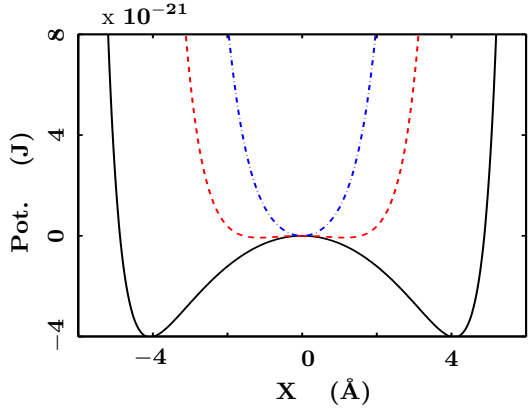}\\
  \vspace{5 mm}
  \caption{\label{fig:potential} The total potential arising from
    elastic and vdW-interactions as a function of the grain position
    $X$ shown for three different values of the ligand stiffness $k$.
    The solid line (black, \mbox{$k=0.001$ Nm$^{-1}$}) correspond to
    region {\bf I} in \mbox{Fig.  \ref{fig:ivk}} while the dashed line
    (red, \mbox{$k=0.05$ Nm$^{-1}$}) and the dash-dotted (blue,
    \mbox{$k=0.3$ Nm$^{-1}$}) correspond to the regions {\bf II} and
    {\bf III} respectively.}
\end{figure}

We have chosen numerical values for the model system with a typical
state of the art experimental nanoelectromechanical system in mind.
We have considered a grain of mass \mbox{$M=5\times 10^{-23}$ kg} and
radius \mbox{$r=1$ nm} placed between two leads separated by a
distance of \mbox{$L=5$ nm}. The elastic potential was chosen to have
a minimum in the center of the system, i.e. $X_0=0$, and to diverge at
a minimum grain-lead separation of \mbox{$0.5$ nm} which is just the
statement that the ligands can never be compressed to less than a
third of their size.  The tunneling resistances and capacitances
corresponding to zero grain displacement were \mbox{$R_0^{L,R} =
  10^{9} \Omega$} and \mbox{$C_0^{L,R} = 10^{-18}$ F} which
corresponds to room temperature Coulomb blockade behavior.  The
tunneling length is material dependent but can be estimated to be of
the order $1$~{\AA}.  The capacitance length scale is determined by
the system size and geometry and is estimated to \mbox{$A_{L,R} = 2.5$
  nm}.  For the vdW-interaction we have used the attractive Hamaker
constant, $H_{attr}=4\times 10^{-19}$J~\cite{a:israelachvili_85} and
$H_{rep}$ was chosen from dimensional considerations, to be
$10^{-72}\textrm{Jm}^6$.  The focus on this work is the behavior of
the shuttle instability as the relative strength of the elastic and
van der Waals interactions change. We thus chose to consider an
interval of the elasticity constant from $k=10^{-6}$~Nm$^{-1}$ to
$k=1$~Nm$^{-1}$ corresponding to small oscillation eigenfrequencies of
the order of $0.1$~GHz to $100$~GHz. This spans the interval of
frequencies typical for NEMS\cite{a:01_roukes}.  The damping constant
used was \mbox{$\gamma = 10^{-13}$ kg s$^{-1}$} corresponding to
quality factors of approximately $0.5 \times 10^3$ through $0.5 \times
10^6$ which are also typical for those observed in
NEMS\cite{a:01_roukes}.  In Fig.~\ref{fig:system2} the vdW-potential
and the elastic potential is plotted together with their sum. The
whole graph is superimposed on a schematic figure of the system
showing the grain and the leads. The dashed curve (red) shows the
elastic potential arising from the the molecular links while the thin
solid line (blue) is the vdW-force curve. The total potential is shown
as a thick line (black). It is clear that if, as is the case in our
model, the insulating material does not allow the grain to come into
direct contact with the leads the repulsive part of the vdW force does
not come into play at all. It is also clear that the relation between
the elastic force and the vdW-force determines the qualitative
behavior of the total potential.  Indeed, if the stiffness of the
ligands is small the potential will have two stable minima whereas a
stiffer configuration may have only one. The total potentials arising
from different stiffnesses are shown in \mbox{Fig.
  \ref{fig:potential}}.

\begin{figure}
  \psfrag{xm6}[]{\hspace*{7mm}\raisebox{0mm}{{$\mathbf{10^{-6}}$}}}
  \psfrag{xm4}[]{\hspace*{7mm}\raisebox{0mm}{{$\mathbf{10^{-4}}$}}}
  \psfrag{xm2}[]{\hspace*{7mm}\raisebox{0mm}{{$\mathbf{10^{-2}}$}}}
  \psfrag{xm0}[]{\hspace*{7mm}\raisebox{0mm}{{$\mathbf{10^{-0}}$}}}
  \psfrag{xlabel}[]{\hspace*{7mm}\raisebox{-3mm}{{\bf k}\hspace*{2mm}
      {\bf (Nm$^{-1}$)}}}
  \psfrag{y0}[]{\hspace*{3mm}\raisebox{2mm}{{$\mathbf{0}$}}}
  \psfrag{y5}[]{\hspace*{4mm}\raisebox{2mm}{{$\mathbf{0.5}$}}}
  \psfrag{y1}[]{\hspace*{4mm}\raisebox{2mm}{{$\mathbf{1}$}}}
  \psfrag{ylabel}[]{\hspace*{10mm}\raisebox{-4mm}{{\bf V}\hspace*{2mm}
      {\bf (V)}}}
  \psfrag{zm12}[]{\hspace*{12mm}\raisebox{4mm}{{$\mathbf{10^{-12}}$}}}
  \psfrag{zm10}[]{\hspace*{12mm}\raisebox{4mm}{{$\mathbf{10^{-10}}$}}}
  \psfrag{zm8}[]{\hspace*{12mm}\raisebox{4mm}{{$\mathbf{10^{-8}}$}}}
  \psfrag{zlabel}[]{\hspace*{5mm}\raisebox{8mm}{\bf{I}\hspace*{2mm}
      {\bf (A)}}} \psfrag{R1}[]{\hspace*{0mm}\raisebox{0mm}{{\bf
        III}}} \psfrag{R2}[]{\hspace*{0mm}\raisebox{0mm}{{\bf II}}}
  \psfrag{R3}[]{\hspace*{0mm}\raisebox{0mm}{{\bf I}}}
  \includegraphics[width=8cm]{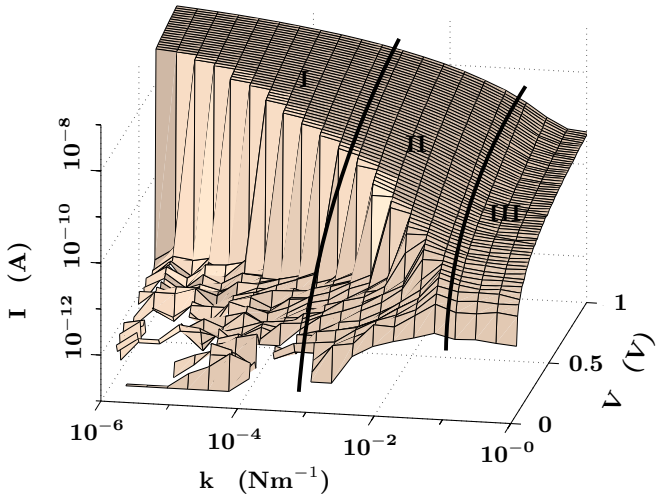} \vspace{5 mm}
  \caption{\label{fig:ivk} The current plotted as a function of
    the bias voltage $V$ and the spring constant $k$. The figure shows
    that the transition changes character from a very sharp transition
    in the region marked {\bf I} to a more smooth transition in the
    region marked {\bf III}. The intermediate region is marked {\bf
      II}. In this graph all IV-curves correspond to increasing
    voltage sweeps which means that hysteresis effects are not shown
    here.}
\end{figure}

In Fig.~\ref{fig:ivk} we have plotted numerical results for the
IV-characteristics as a function of the elasticity constant $k$. The
figure is divided into three regions. In region {\bf I}, the
elasticity, $k$, is small and the potential has two stable minima; one
in the vicinity of each lead. In region {\bf III}, $k$ is large enough
for the potential to have a single minimum in the center of the
system. The potentials in \mbox{Fig. \ref{fig:potential}} are
representative of the different regions. The IV-characteristics in
these regions differ markedly. Below we will elucidate this behavior
with a mainly qualitative discussion of the dynamics, starting with
the case when $k$ is small, i.e. region {\bf I}. Then moving on to
discuss region {\bf III} and finally we comment on the intermediate
region, labelled region {\bf II} in Fig.~\ref{fig:ivk}.

\subsection{Small elasticity, region {\bf I}}
\begin{figure}
  \psfrag{xi03}[]{\hspace*{3mm}\raisebox{0mm}{\tiny$\mathbf{0.3}$}}
  \psfrag{xi04}[]{\hspace*{3mm}\raisebox{0mm}{\tiny$\mathbf{0.4}$}}
  \psfrag{yi1}[]{\hspace*{3mm}\raisebox{1mm}{\tiny$\mathbf{1}$}}
  \psfrag{yi3}[]{\hspace*{3mm}\raisebox{1mm}{\tiny$\mathbf{3}$}}
  \psfrag{yi5}[]{\hspace*{3mm}\raisebox{1mm}{\tiny$\mathbf{5}$}}
  \psfrag{yiexp}[]{\hspace*{9mm}\raisebox{3mm}{\tiny x $\mathbf{10^{-9}}$}}
  \psfrag{xilabel}[]{\hspace*{4mm}\raisebox{-3mm}{{\bf \tiny V}\hspace*{2mm}
      {\bf \tiny (V)}}}
  \psfrag{yilabel}[]{\hspace*{6mm}\raisebox{0mm}{{\bf \tiny I}\hspace*{2mm}
      {\bf \tiny (A)}}}
  \psfrag{aa}[]{\hspace*{-10mm}\raisebox{0mm}{(a)}}
  \psfrag{bb}[]{\hspace*{-10mm}\raisebox{4mm}{(b)}}
  \psfrag{cc}[]{\hspace*{-10mm}\raisebox{4mm}{(c)}}
  \psfrag{ayexp}[]{\hspace*{15mm}\raisebox{4mm}{x $\mathbf{10^{-8}}$}}
  \psfrag{axnoll}[]{\hspace*{7mm}\raisebox{0mm}{$\mathbf{0}$}}
  \psfrag{ax02}[]{\hspace*{5mm}\raisebox{0mm}{$\mathbf{0.2}$}}
  \psfrag{ax04}[]{\hspace*{5mm}\raisebox{0mm}{$\mathbf{0.4}$}}
  \psfrag{ax06}[]{\hspace*{5mm}\raisebox{0mm}{$\mathbf{0.6}$}}
  \psfrag{ax08}[]{\hspace*{5mm}\raisebox{0mm}{$\mathbf{0.8}$}}
  \psfrag{ax10}[]{\hspace*{5mm}\raisebox{0mm}{$\mathbf{1}$}}
  \psfrag{axlabel}[]{\hspace*{15mm}\raisebox{0mm}{{\bf V}\hspace*{2mm}
      {\bf (V)}}}
  \psfrag{ay10}[]{\hspace*{3mm}\raisebox{2mm}{$\mathbf{1.0}$}}
  \psfrag{ay20}[]{\hspace*{3mm}\raisebox{2mm}{$\mathbf{2.0}$}}
  \psfrag{aylabel}[]{\hspace*{7mm}\raisebox{2mm}{{\bf I}\hspace*{2mm}
      {\bf (A)}}}
  \psfrag{by00}[]{\hspace*{4mm}\raisebox{3mm}{}}
  \psfrag{by06}[]{\hspace*{7mm}\raisebox{3mm}{\small$-0.075$}}
  \psfrag{bym06}[]{\hspace*{-2mm}\raisebox{3mm}{\small$-0.08$}}
  \psfrag{cx00}[]{\hspace*{5mm}\raisebox{0mm}{\small$0$}}
  \psfrag{cx10}[]{\hspace*{5mm}\raisebox{0mm}{\small$5$}}
  \psfrag{cx20}[]{\hspace*{5mm}\raisebox{0mm}{\small$10$}}
  \psfrag{bylabel}[]{\hspace*{8mm}\raisebox{2mm}{\small {\bf X/L}}}
  \psfrag{bxlabel}[]{\hspace*{14mm}\raisebox{-4mm}{\small {\bf t}\hspace*{2mm}
      {\bf (ns)}}}
  \psfrag{cylabel}[]{\hspace*{8mm}\raisebox{2mm}{\small {\bf X/L}}}
  \psfrag{cxlabel}[]{\hspace*{14mm}\raisebox{-4mm}{\small {\bf t}\hspace*{2mm}
      {\bf (ns)}}}
  \psfrag{cy00}[]{\hspace*{4mm}\raisebox{3mm}{\small$0$}}
  \psfrag{cy10}[]{\hspace*{4mm}\raisebox{3mm}{\small$0.1$}}
  \psfrag{cym10}[]{\hspace*{2mm}\raisebox{3mm}{\small$-0.1$}}
  \includegraphics[width=8cm, height=10cm]{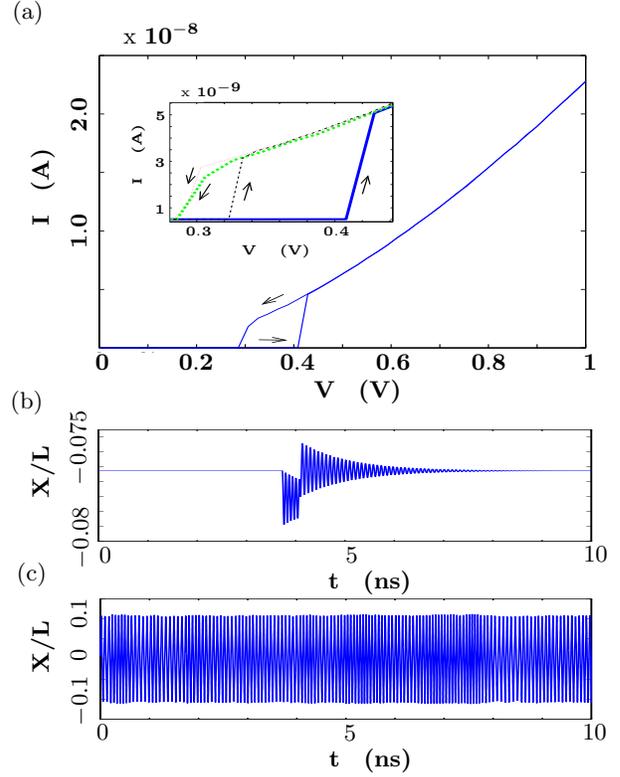}
  \vspace{5 mm}
  \caption{\label{fig:hyst_low} (a) IV-characteristics for
    \mbox{$k = 0.001$ Nm$^{-1}$} at zero temperature. The inset shows
    a closeup of the transition region. Here IV-curves for $T=0$~K for
    increasing (blue, thick solid) and decreasing (green, thick
    dashed) voltages are shown together with IV-curves for $T=300$~K
    for increasing (black, thin dash-dotted) and decreasing (red, thin
    dotted) voltages. (b) Position of the grain as a function of time
    for \mbox{$V=0.36$ V} in the case when the voltage is swept up.
    The grain sits asymmetrically in one of the potential minima
    caused by the strong van der Waals forces. (c) Position of the
    grain as a function of time for \mbox{$V=0.36$ V} in the case when
    the voltage is swept down.  Since the grain is already oscillating
    it does not get stuck in the potential minimum and the current is
    still large.}
\end{figure}

In \mbox{Fig. \ref{fig:hyst_low}a} we have plotted the
IV-characteristics for \mbox{$k=0.001$ Nm$^{-1}$}. The most important
characteristic is the big hysteresis loop in the IV-curve. The two
lower panels (b and c) of \mbox{Fig. \ref{fig:hyst_low}} show the
grain position as a function of time for low and high voltages
respectively.
For bias voltages around zero the the grain will settle in one of the
two potential minima. Since the total current through the system is
limited by the larger of the two tunnel resistances it will be
strongly (exponentially) suppressed as compared to the situation when
the grain occupies a more central position. As the voltage is
increased the electrostatic force on the grain changes. The reason for
this change is twofold; first, the electric field strength in the
inter-electrode gap increases, and, second, increasing the bias will
allow more charge to reside on the grain. When the voltage is further
increased it eventually reaches a point $V=V_{\rm u}$ where the grain
can escape its locked position. In \mbox{Fig. \ref{fig:hyst_low}a}
this happens at a bias voltage just above $0.4$~V. Below this bias the
grain cannot escape its locked asymmetric position, even though, as
shown in \mbox{Fig.  \ref{fig:hyst_low}b}, tunneling events can occur
causing small movements of the grain around the local potential
minimum. Above $V=V_{\rm u}$, self oscillations of the grain appears,
which causes an increase in the current by several orders of
magnitude.

\begin{figure}
  \psfrag{p1}[]{\hspace*{4mm}\raisebox{0mm}{(1)}}
  \psfrag{p2}[]{\hspace*{2mm}\raisebox{-2mm}{(2)}}
  \psfrag{p3}[]{\hspace*{-6mm}\raisebox{4mm}{(3)}}
  \psfrag{p4}[]{\hspace*{-10mm}\raisebox{1mm}{(4)}}
  \psfrag{n3}[]{\hspace*{14mm}\raisebox{-5mm}{$n=3$}}
  \psfrag{n4}[]{\hspace*{24mm}\raisebox{-10mm}{$n=4$}}
  \psfrag{xlabel}[]{\hspace*{10mm}\raisebox{-5mm}{{\bf X} \hspace*{2mm}
  {\bf (nm)}}}
  \psfrag{ylabel}[]{\hspace*{3mm}\raisebox{2mm}{{\bf Pot.} \hspace*{2mm}
  {\bf (arb. units)}}}
  \psfrag{xexp}[]{\hspace*{0mm}\raisebox{0mm}{\mbox{}}}
  \psfrag{x0}[]{\hspace*{2mm}\raisebox{-2mm}{$\mathbf{0}$}}
  \psfrag{xm2}[]{\hspace*{0mm}\raisebox{-2mm}{$\mathbf{-0.2}$}}
  \psfrag{xm4}[]{\hspace*{0mm}\raisebox{-2mm}{$\mathbf{-0.4}$}}
  \includegraphics[width=7cm, height=5cm]{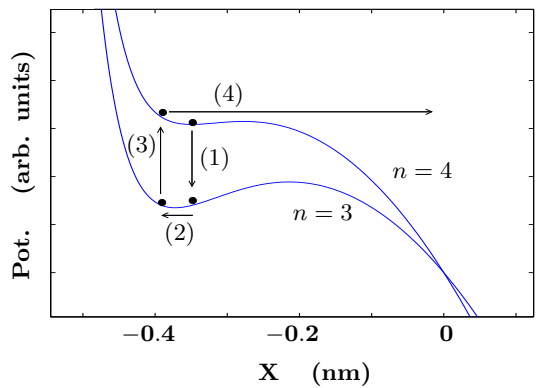}
  \vspace{5 mm}
  \caption{\label{fig:escape_illustration} Example of grain escape by
    a single pumping cycle. A grain initially carrying $n=4$ extra
    electrons and at rest in the local minimum of the total potential
    cannot escape over the barrier at low temperatures. However, a
    charging and motion sequence as described by the arrows (1)
    through (4) provides the necessary energy to overcome the
    barrier.}
\end{figure}

The transition from the locked position can occur in several different
ways. In the simplest case, in the absence of thermal over-barrier
activation and when the dissipation is large, the pumping due to grain
motion inside the local minimum is very small, and the grain cannot
escape until the minimum vanishes. However, if the minimum is shallow
and the dissipation is not too high, correlation of charging and
motion of the grain in the local potential minimum can lead to energy
pumping and grain escape. An example of such a sequence of correlated
events is illustrated in \mbox{Fig. \ref{fig:escape_illustration}}.
The figure illustrates a grain initially having $n=4$ extra electrons
and that is positioned in the local minimum of the total potential.
The numbered arrows correspond to the following events: (1) A
tunneling event occurs changing the charge from $n=4$ to $n=3$. This
change in charge is associated with a corresponding change in the
potential.  The grain is now not positioned in the minimum of the
$n=3$ potential. (2) The grain moves in the $n=3$ potential until a
new tunneling event occurs. (3) A tunneling event changes the charge
back to $n=4$. (4) The grain, again in the potential for $n=4$, now
has enough energy to escape the potential well.

The main IV-curve in Fig.~\ref{fig:hyst_low}a corresponds to the case
of zero temperature.  In this case, the maximum charge allowed on the
grain, and hence, the most "shallow potential" experienced by the
grain, is determined by the bias voltage due to charging effects
(Coulomb blockade). At finite temperatures, thermal fluctuations of
the grain charge can cause the grain to become released at a lower
voltage. In the inset of Fig.~\ref{fig:hyst_low}a a close up of the
transition region for both a finite ($T=300$~K) and zero temperature
is shown. As can be seen, the threshold were the grain escapes its
locked position is considerably lowered for the high temperature case,
while the remaining part of the curve remains unchanged.

\begin{figure}
  \psfrag{a}[]{\hspace*{10mm}\raisebox{0mm}{(a)}}
  \psfrag{b}[]{\hspace*{10mm}\raisebox{0mm}{(b)}}
  \psfrag{ayexp}[]{\hspace*{17mm}\raisebox{3mm}{x $\mathbf{10^{-8}}$}}
  \psfrag{axnoll}[]{\hspace*{0mm}\raisebox{-2.7mm}{$\mathbf{0}$}}
  \psfrag{0.2}[]{\hspace*{2mm}\raisebox{-4mm}{$\mathbf{0.2}$}}
  \psfrag{0.4}[]{\hspace*{2mm}\raisebox{-4mm}{$\mathbf{0.4}$}}
  \psfrag{0.6}[]{\hspace*{2mm}\raisebox{-4mm}{$\mathbf{0.6}$}}
  \psfrag{xlabel}[]{\hspace*{5mm}\raisebox{-10mm}{{\bf V}\hspace*{2mm}
      {\bf (V)}}}
  \psfrag{ay1}[]{\hspace*{4mm}\raisebox{2.5mm}{$\mathbf{1.0}$}}
  \psfrag{ay2}[]{\hspace*{4mm}\raisebox{2.5mm}{$\mathbf{2.0}$}}
  \psfrag{ylabel}[]{\hspace*{5mm}\raisebox{10mm}{{\bf I}\hspace*{2mm}
      {\bf (A)}}}
  \psfrag{byexp}[]{\hspace*{17mm}\raisebox{3mm}{x $\mathbf{10^{-9}}$}}
  \psfrag{bxnoll}[]{\hspace*{0mm}\raisebox{-2.5mm}{$\mathbf{0}$}}
  \psfrag{by1}[]{\hspace*{4mm}\raisebox{2mm}{$\mathbf{1.0}$}}
  \psfrag{by2}[]{\hspace*{3mm}\raisebox{2mm}{$\mathbf{2.0}$}}
  \psfrag{by3}[]{\hspace*{4mm}\raisebox{2mm}{$\mathbf{3.0}$}}
  \psfrag{xnamn}[]{\hspace*{5mm}\raisebox{-10mm}{{\bf V}\hspace*{2mm}
      {\bf (V)}}}
  \psfrag{ynamn}[]{\hspace*{5mm}\raisebox{10mm}{{\bf I}\hspace*{2mm}
      {\bf (A)}}}
  \includegraphics[width=8cm, height=10cm]{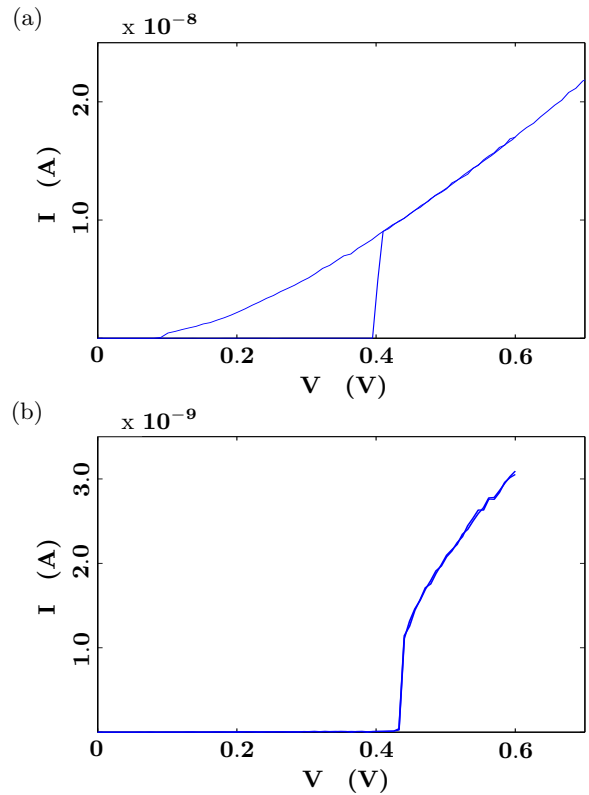}
  \vspace{5 mm}
  \caption{\label{fig:hyst_test} The IV-characteristics of the
    same system as described in \mbox{Fig. \ref{fig:hyst_low}} but
    with different dissipation factors $\gamma$ to investigate
    features of the hysteresis. In (a) the dissipation is lowered to
    $\gamma = 10^{-14}$ kg s$^{-1}$ and we see that the hysteresis is
    bigger than in \mbox{Fig. \ref{fig:hyst_low}}. In (b) the
    dissipation is instead increased to $\gamma = 10^{-12}$ kg
    s$^{-1}$ and the hysteresis has disappeared completely.}
\end{figure}

To understand the hysteresis appearing in the curve consider a voltage
above $V_{\rm u}$. Provided that the dissipation is not too high a
sustained oscillation of the grain between the leads results. Whether
or not this oscillation occurs depends, once the grain is released, on
the relation between the work done by the electric field between the
leads and the dissipation in the
system~\cite{a:98_gorelik,a:98_isacsson}. Lowering the voltage below
$V_{\rm u}$ once the grain is oscillating does then not imply that it
will be stuck in one of the local minima of the potential. Hence, by
changing the magnitude of the dissipation the size of the hysteresis
loop varies. This is illustrated in Fig.~\ref{fig:hyst_test} where a
case with large hysteresis ($\gamma=10^{-14}$ kg s$^{-1}$) and a case
with no hysteresis ($\gamma=10^{-12}$ kg s$^{-1}$) are shown. It
should be noted that the current in Fig.~\ref{fig:hyst_test}a and b
differ by one order of magnitude. The dynamics of the latter case,
when dissipation dominates the behavior, is mainly the same as in
Ref.~\cite{a:02_nord} where it was further investigated.

\subsection{Large elasticity, region {\bf III}}
\begin{figure}
  \psfrag{aa}[]{\hspace*{-10mm}\raisebox{0mm}{(a)}}
  \psfrag{bb}[]{\hspace*{-10mm}\raisebox{4mm}{(b)}}
  \psfrag{cc}[]{\hspace*{-10mm}\raisebox{4mm}{(c)}}
  \psfrag{ayexp}[]{\hspace*{15mm}\raisebox{4mm}{x $\mathbf{10^{-9}}$}}
  \psfrag{axnoll}[]{\hspace*{7mm}\raisebox{0mm}{$\mathbf{0}$}}
  \psfrag{ax02}[]{\hspace*{5mm}\raisebox{0mm}{$\mathbf{0.2}$}}
  \psfrag{ax04}[]{\hspace*{5mm}\raisebox{0mm}{$\mathbf{0.4}$}}
  \psfrag{ax06}[]{\hspace*{5mm}\raisebox{0mm}{$\mathbf{0.6}$}}
  \psfrag{ax08}[]{\hspace*{5mm}\raisebox{0mm}{$\mathbf{0.8}$}}
  \psfrag{ax10}[]{\hspace*{5mm}\raisebox{0mm}{$\mathbf{1}$}}
  \psfrag{axlabel}[]{\hspace*{15mm}\raisebox{0mm}{{\bf V}\hspace*{2mm}
      {\bf (V)}}}
  \psfrag{ay10}[]{\hspace*{3mm}\raisebox{2mm}{$\mathbf{0.4}$}}
  \psfrag{ay20}[]{\hspace*{3mm}\raisebox{2mm}{$\mathbf{0.8}$}}
  \psfrag{aylabel}[]{\hspace*{7mm}\raisebox{2mm}{{\bf I}\hspace*{2mm}
      {\bf (A)}}}
  \psfrag{by00}[]{\hspace*{1mm}\raisebox{3mm}{\small $0$}}
  \psfrag{by04}[]{\hspace*{4mm}\raisebox{3mm}{\small $0.004$}}
  \psfrag{bym04}[]{\hspace*{-4mm}\raisebox{3mm}{\small $-0.004$}}
  \psfrag{cx00}[]{\hspace*{5mm}\raisebox{0mm}{\small $0$}}
  \psfrag{cx10}[]{\hspace*{5mm}\raisebox{0mm}{\small $5$}}
  \psfrag{cx20}[]{\hspace*{5mm}\raisebox{0mm}{\small $10$}}
  \psfrag{bylabel}[]{\hspace*{8mm}\raisebox{2mm}{\small {\bf X/L}}}
  \psfrag{bxlabel}[]{\hspace*{14mm}\raisebox{0mm}{\small {\bf t}\hspace*{2mm}
      {\bf (ns)}}}
  \psfrag{cylabel}[]{\hspace*{8mm}\raisebox{2mm}{\small {\bf X/L}}}
  \psfrag{cxlabel}[]{\hspace*{14mm}\raisebox{0mm}{\small {\bf t}\hspace*{2mm}
      {\bf (ns)}}}
  \psfrag{cy00}[]{\hspace*{4mm}\raisebox{3mm}{\small $0$}}
  \psfrag{cy10}[]{\hspace*{6mm}\raisebox{3mm}{\small $0.03$}}
  \psfrag{cym10}[]{\hspace*{1mm}\raisebox{3mm}{\small $-0.03$}}
  \includegraphics[width=8cm, height=10cm]{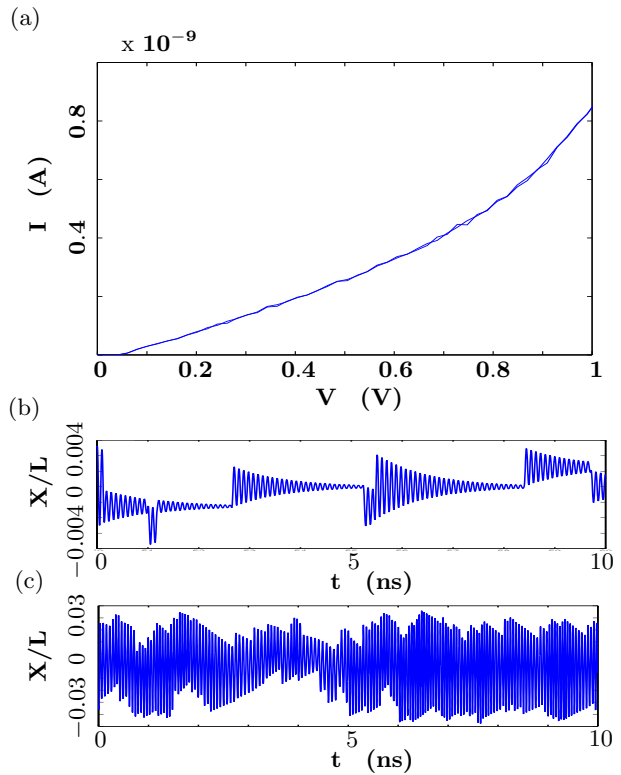}
  \vspace{5 mm}
  \caption{\label{fig:hyst_high} (a) IV-curve for \mbox{$k = 0.3$
      Nm$^{-1}$}. (b) Position of the grain as a function of time for
    \mbox{$V=0.2$ V}. The grain oscillates in the single minimum in
    the potential. The position of the minimum changes with the number
    of charges on the grain due to the applied electric field.  (c)
    Position of the grain as a function of time for \mbox{$V=1.0$ V}.
    As the voltage is increased the grain starts to oscillate across
    the system.}
\end{figure}

In the limit when the elasticity dominates the short range attractive
forces only one minimum results from the elastic and vdW-potentials.
This yields IV-characteristics belonging to region {\bf III} in
Fig.~\ref{fig:ivk}. The particular curve corresponding to $k=0.3$
Nm$^{-1}$ is shown in \mbox{Fig.~\ref{fig:hyst_high}a}. In this case,
as the bias voltage is increased the shuttle instability develops from
the one minimum in the potential. Sweeping the voltage down again
reveals the absence of hysteresis.

It should be noted here that, although the system operates in a regime
analogous to the shuttle system in Ref.~\cite{a:98_gorelik}, no steps
are visible in the IV-curve even at zero temperature as would be the
case if the elastic potential was purely harmonic. For a harmonic
potential, increasing the bias leads to larger oscillation amplitudes
at constant frequency, meaning that the grain comes closer to the
lead. Since the tunnel resistance decreases exponentially with
decreasing grain-lead distance the tunnel rate increases
exponentially. Thus, increasing the amplitude at constant oscillation
frequency implies that the grain charge has time to reach complete
equilibrium with the lead. This in turn gives maximum step sharpness
in the IV-characteristics~\cite{a:99_weiss}. The presence of the
strong anharmonicity though, confining the grain to the center of the
system, implies that with increasing bias the energy pumped into the
vibrational degrees of freedom increases not only the amplitude but
also the frequency. When the frequency increases the time spent by the
grain in close contact with the lead decreases. Unless the amplitude
is increased accordingly the grain charge does not have time to reach
equilibrium with the lead, leading to the suppression of the steps seen
in the IV-curves presented here.

Although it is hard to tell from the curve in
\mbox{Fig.~\ref{fig:hyst_high}a} that shuttling is present, it is more
obvious from \mbox{Fig.~\ref{fig:hyst_high}b} and \mbox{Fig.
  \ref{fig:hyst_high}c} where the position of the grain as a function
of time is shown for the voltages \mbox{$V=0.2$ V} and \mbox{$V=1.0$ V}.

\subsection{Intermediate elasticity, region {\bf II}}
\begin{figure}
  \psfrag{aa}[]{\hspace*{-10mm}\raisebox{0mm}{(a)}}
  \psfrag{bb}[]{\hspace*{-10mm}\raisebox{4mm}{(b)}}
  \psfrag{cc}[]{\hspace*{-10mm}\raisebox{4mm}{(c)}}
  \psfrag{ayexp}[]{\hspace*{15mm}\raisebox{3mm}{x $\mathbf{10^{-9}}$}}
  \psfrag{axnoll}[]{\hspace*{7mm}\raisebox{0mm}{$\mathbf{0}$}}
  \psfrag{ax02}[]{\hspace*{5mm}\raisebox{0mm}{$\mathbf{0.2}$}}
  \psfrag{ax04}[]{\hspace*{5mm}\raisebox{0mm}{$\mathbf{0.4}$}}
  \psfrag{ax06}[]{\hspace*{5mm}\raisebox{0mm}{$\mathbf{0.6}$}}
  \psfrag{ax08}[]{\hspace*{5mm}\raisebox{0mm}{$\mathbf{0.8}$}}
  \psfrag{ax10}[]{\hspace*{5mm}\raisebox{0mm}{$\mathbf{1}$}}
  \psfrag{axlabel}[]{\hspace*{15mm}\raisebox{0mm}{{\bf V}\hspace*{2mm}
      {\bf (V)}}}
  \psfrag{ay10}[]{\hspace*{3mm}\raisebox{2mm}{$\mathbf{1.0}$}}
  \psfrag{ay20}[]{\hspace*{3mm}\raisebox{2mm}{$\mathbf{2.0}$}}
  \psfrag{ay30}[]{\hspace*{3mm}\raisebox{2mm}{$\mathbf{3.0}$}}
  \psfrag{aylabel}[]{\hspace*{7mm}\raisebox{2mm}{{\bf I}\hspace*{2mm}
      {\bf (A)}}}
  \psfrag{by00}[]{\hspace*{2mm}\raisebox{3mm}{\small$0$}}
  \psfrag{by02}[]{\hspace*{7mm}\raisebox{3mm}{\small$0.02$}}
  \psfrag{bym02}[]{\hspace*{-1mm}\raisebox{3mm}{\small$-0.02$}}
  \psfrag{cx00}[]{\hspace*{5mm}\raisebox{0mm}{\small $0$}}
  \psfrag{cx10}[]{\hspace*{5mm}\raisebox{0mm}{\small $5$}}
  \psfrag{cx20}[]{\hspace*{5mm}\raisebox{0mm}{\small $10$}}
  \psfrag{bylabel}[]{\hspace*{8mm}\raisebox{2mm}{\small {\bf X/L}}}
  \psfrag{bxlabel}[]{\hspace*{14mm}\raisebox{0mm}{\small {\bf t}\hspace*{2mm}
      {\bf (ns)}}}
  \psfrag{cylabel}[]{\hspace*{8mm}\raisebox{2mm}{\small {\bf X/L}}}
  \psfrag{cxlabel}[]{\hspace*{14mm}\raisebox{0mm}{\small {\bf t}\hspace*{2mm}
      {\bf (ns)}}}
  \psfrag{cy00}[]{\hspace*{3mm}\raisebox{3mm}{\small$0$}}
  \psfrag{cy10}[]{\hspace*{4mm}\raisebox{3mm}{\small$0.04$}}
  \psfrag{cym10}[]{\hspace*{-1mm}\raisebox{3mm}{\small$-0.04$}}
  \includegraphics[width=8cm, height=10cm]{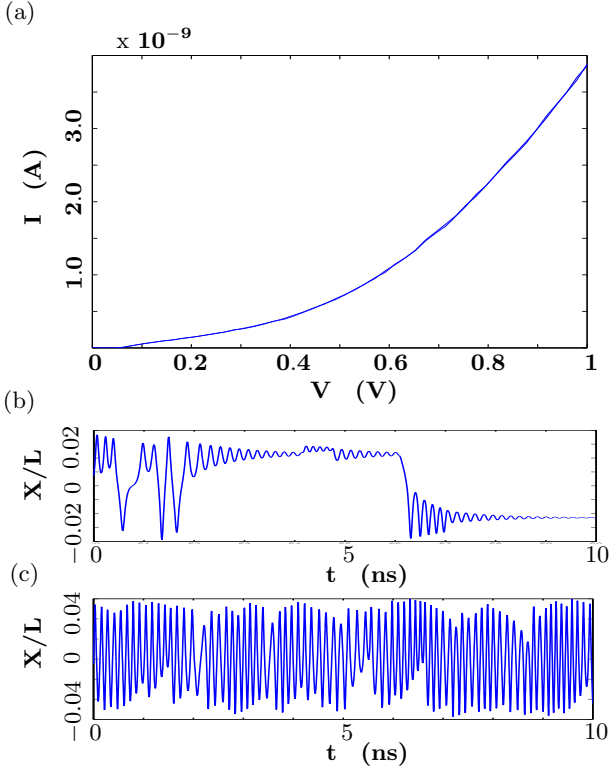}
  \vspace{5 mm}
  \caption{\label{fig:hyst_inter} (a) IV-characteristics for \mbox{$k =
      0.05$ Nm$^{-1}$}. (b) Position of the grain as a function of
    time for \mbox{$V=0.1$ V}. For low voltages the grain oscillates
    in one potential minimum until a tunnel event changes its charge.
    It is then is moved across the system to the other minimum where
    it oscillates until the charge changes again.  (c) Position of the
    grain as a function of time for \mbox{$V=0.4$ V}. For higher
    voltages the grain starts to oscillate across the system.}
\end{figure}

In the intermediate region the potential minima are not as deep as in
region {\bf I} (cf. Fig.~\ref{fig:potential}). An interesting
phenomenon can occur in this regime; As the bias voltage is raised the
grain can sit in one minimum, then be pushed to the other, oscillate
there for a short time, and then change side again.  This behavior is
shown in Fig.~\ref{fig:hyst_inter}b where the position of the grain as
a function of time is plotted for \mbox{$k=0.05$ Nm$^{-1}$} and
\mbox{$V=0.1$ V}.

Raising the bias, the grain switches side with increasing frequency
(see Fig.~\ref{fig:hyst_inter}c) until it changes side twice in each
oscillation period. Since the potential is flatter in this regime, the
shuttle transition occurs at a lower voltage than in region {\bf III}.

\section{Conclusions}
We have investigated the effects of van der Waals forces on the
current transport through a shuttle system. Our results show that the
relative strength between these forces changes how the transition from
non-shuttle to shuttle transport occurs. In the region where the
elastic force is small the transition is very sharp and there is a big
hysteresis loop in the IV-curve, whereas in the region where the
elastic force is big the transition is softer and there is no
hysteresis. Between the two regions is a crossover region where the
transition voltage is significantly lowered and where, for lower
voltages, the grain alternatingly oscillates inside and between the
minima of the potential.

\section{Acknowledgment}
We would like to thank Robert Shekhter and Leonid Gorelik for valuable
discussions and useful remarks on this manuscript. One of the authors
(A.I) acknowledge financial support from the Swedish SSF through the
programme QDNS and from the Swedish Foundation for International
Cooperation in Research and Higher Education (STINT). The other author
(T.N.) acknowledge financial support from the Swedish Research Council
(VR).

\bibliographystyle{plain}

\begin{thebibliography}{99}

\bibitem{a:01_roukes}M. Roukes, Phys. World \textbf{14}(2), 25
  (2001); Opening Lecture, 2000, Solid State Sensor and Actuator Workshop,
Hilton Head, SC 6/4/2000, published in Technical Digest of the 2000 Solid State
Sensor and Actuator Workshop.

\bibitem{a:99_kim}P. Kim and C. M. Lieber, Science {\bf 286}, 2148
  (1999).

\bibitem{a:00_blick}R.H. Blick, A. Erbe, H. Kr{\"o}mmer, A. Kraus, and
  J.P. Kotthaus, Physica E \textbf{6}, 821 (2000).

\bibitem{a:00_rueckes}T. Rueckes, K. Kim, E. Joselevich, G.Y. Tseng, C. Cheung,
  and C.M. Lieber, Science \textbf{289}, 94 (2000).

\bibitem{a:03_kinaret}J.M. Kinaret, T. Nord, and S. Viefers,
  Appl. Phys. Lett. \textbf{82}, 1287 (2003).

\bibitem{a:98_gorelik}L.Y. Gorelik, A. Isacsson, M.V. Voinova,
  B. Kasemo, R.I. Shekhter, and M. Jonson, Phys. Rev. Lett. \textbf{80},
  4526 (1998).

\bibitem{a:98_erbe}A. Erbe, R.H. Blick, A. Tilke, A. Kriele, and
  J.P. Kotthaus, Appl. Phys. Lett. \textbf{73}, 3751 (1998).

\bibitem{a:01_erbe}A. Erbe, C. Weiss, W. Zwerger, and R.H. Blick,
  Phys. Rev. Lett. \textbf{87}, 096106 (2001).

\bibitem{a:02_scheible}D. V. Scheible, A. Erbe, and R.H. Blick, New
  J. Phys. \textbf{4}, 86 (2002).

\bibitem{a:00_park}H. Park, J. Park, A.K.L Lim, E.H. Anderson,
  A.P. Alivisatos, and P.L. McEuen, Nature (London) \textbf{407}, 57
  (2000).

\bibitem{a:02_nagano}K. Nagano, A. Okuda, and Y. Majima,
  Appl. Phys. Lett. \textbf{81}, 544 (2002).

\bibitem{a:99_tuominen}M.T. Tuominen, R.V. Krotkov, and M.L. Breuer,
  Phys. Rev. Lett. \textbf{83}, 3025 (1999).

\bibitem{a:98_isacsson}A. Isacsson, L.Y. Gorelik, M.V. Voinova,
  B. Kasemo, R.I. Shekhter, and M. Jonson, Physica B \textbf{255}, 150
  (1998).

\bibitem{a:01_isacsson}A. Isacsson, Phys. Rev. B \textbf{64}, 035326
  (2001).

\bibitem{a:01_nishiguchi}N. Nishiguchi, Phys. Rev. B \textbf{65},
  035403 (2001).

\bibitem{a:02_nord}T. Nord, L.Y. Gorelik, R.I. Shekhter, and
  M. Jonson, Phys. Rev. B \textbf{65}, 165312 (2002).

\bibitem{a:99_weiss}C. Weiss, W. Zwerger, Europhys. Lett. \textbf{47},
  97 (1999).

\bibitem{a:02_nishiguchi}N. Nishiguchi, Phys. Rev. Lett. \textbf{89},
  66802 (2002).

\bibitem{a:01_boese}D. Boese, H. Schoeller,
  Europhys. Lett. \textbf{54}, 668 (2001).

\bibitem{a:02_fedorets1}D. Fedorets, L.Y. Gorelik, R.I. Shekhter and
  M. Jonson, Europhys. Lett. \textbf{58}, 99 (2002).


\bibitem{a:02_armour}A.D. Armour, and A. MacKinnon, Phys. Rev. B
  \textbf{66}, 035333 (2002).

\bibitem{a:02_fedorets2}D. Fedorets, Phys. Rev. B \textbf{68}, 033106
  (2003).

\bibitem{a:03_novotny}T. Novotn{\'{y}}, A. Donarini, and A.-P. Jauho,
Phys. Rev. Lett. \textbf{90}, 256801 (2003).

\bibitem{a:03_mccarthy}K.D. McCarthy, N. Prokof'ev, and
  M. T. Tuominen, Phys. Rev. B \textbf{67}, 245415 (2003).

\bibitem{a:01_gorelik}L.Y. Gorelik, A. Isacsson, Y.M. Galperin,
  R.I. Shekhter, and M. Jonson, Nature (London) \textbf{411}, 454
  (2001).

\bibitem{a:02_isacsson}A. Isacsson, L. Y. Gorelik, R. I. Shekhter,
  Y. M. Galperin, and M. Jonson, Phys. Rev. Lett. \textbf{89},
  277002 (2002).

\bibitem{a:03_shekhter}R.I. Shekhter, Y. Galperin, L.Y. Gorelik,
  A. Isacsson, and M. Jonson, J. Phys.: Condens. Matter \textbf{15},
  441 (2003).

\bibitem{a:75_kulik}I.O. Kulik and R.I. Shekhter, Sov. Phys. JETP
  \textbf{41}, 308 (1975).

\bibitem{a:averin_91}D.V. Averin and K.K. Likharev, in
  \textit{Mesoscopic Phenomena in Solids}, edited by B.L. Altshuler,
  P.A. Lee, and R.A. Webb (Elsevier, Amsterdam, 1991), p. 173.

\bibitem{a:israelachvili_85} J.N. Israelachvili,
    {\em Intermolecular and Surface Forces}
    (Academic Press, London, 1985).

\end{thebibliography}

\end{document}